\documentclass[aps,prd,twocolumn,showpacs,groupedaddress,nofootinbib,amsmath]{revtex4}
\usepackage{graphicx}
\usepackage{epsfig}
\usepackage{multirow}
\newcommand{\be}{\begin{equation}}
\newcommand{\ee}{\end{equation}}
\newcommand{\ba}{\begin{eqnarray}}
\newcommand{\ea}{\end{eqnarray}}

\begin{document}

\title{Locations of Roberge-Weiss transition endpoints  in lattice QCD with $N_f=2$ improved KS quarks }

\author{Liang-Kai Wu}
\thanks{Corresponding author. Email address: wuliangkai@163.com}
\affiliation{Faculty of Science, Jiangsu University, Zhenjiang 212013, People¡¯s Republic of China}

\author{Xiang-Fei Meng}
\affiliation{National Supercomputer Center in Tianjin, Tianjin, 300457, People¡¯s Republic of China}

\date{\today}

\begin{abstract}
Result on the locations of the tricritical points of $N_f=2$ lattice QCD with
imaginary chemical potential is presented. Simulations are carried out with
Symanzik improved gauge action and  Asqtad  fermion action.  With imaginary chemical potential $i\mu_I=i\pi T$,
previous studies show that the Roberge-Weiss (RW) transition endpoints  are triple points at both large and small quark masses,  and
second order transition points at intermediate quark masses. The
triple and second order endpoints are separated by two tricritical
ones.   Our simulations
are carried  out at 7 values of quark mass $am$
ranging from 0.024 to 0.070 on  lattice volume $12^3\times 4, 16^3\times 4,\, 20^3\times4$. The susceptibility  and Binder
cumulant  of the imaginary part of Polyakov loop are employed to determine the nature
of RW transition endpoints. The simulations suggest  that  the two
tricritical points are within the range $0.024-0.026$ and
$0.040-0.050$, respectively.
\end{abstract}


\maketitle

\section{INTRODUCTION}
\label{SectionIntro}

The phase diagram of QCD has significantly phenomenological implications. It is relevant to the early Universe, compact stars and heavy ion collision experiments.  Reviews on the study of phase diagram can be found in Refs.~\cite{Fukushima:2010bq,Fukushima:2011jc}
 and references therein. While substantial lattice simulation has focused on the phase of QCD at finite density,
a great amount of study centres around QCD with imaginary chemical potential. QCD with  imaginary chemical potential
 has a rich phase structure, and it not only deserves detailed investigations in its
own right theoretically, but also has significant relevance to
physics at zero or small real chemical
potential~\cite{D'Elia:2009qz,D'Elia:2007ke,Bonati:2010gi,Kouno:2009bm,Sakai:2009vb,deForcrand:2010he,Aarts:2010ky,Philipsen:2010rq,Bonati:2012pe}.

The Z(3) symmetry which is present in the pure gauge theory is explicitly broken at the
presence of dynamical quarks. However, Ref.~\cite{Roberge:1986mm} shows that
the Z(3) symmetry is restored when imaginary chemical potential is turned on and
Z(3) transformation can be compensated by a shift in $\mu_I/T $ by
$2\pi/3$, so the partition function of QCD with imaginary  chemical potential has
periodicity in $\mu_I/T $  with period $2\pi/N_c$ as well as
 reflection symmetry in $\mu=i\mu_I$.

Different Z(3) sectors are distinguished  by the phase of
Polyakov loop.  At high temperature, the spontaneous breaking of
Z(3) symmetry  implies transition between adjacent Z(3) sectors in
$\mu_I$ and this transition is of first order,  while at low
temperature, unbroken Z(3) symmetry guarantees  the transition is analytic. The first order transition
takes place at those critical values of imaginary chemical potential
$\mu_I/T =
(2n+1){\pi}/3$~\cite{Roberge:1986mm,deForcrand:2002ci,Lombardo}.  At
high temperature, those first order transition points  form a
transition line which necessarily ends at an endpoint $T_{RW}$ when
the temperature is decreased sufficiently low.

Recent numerical studies~\cite{deForcrand:2010he,D'Elia:2009qz,Bonati:2010gi,Philipsen:2014rpa,Wu:2014lsa} show that the RW transition endpoints
are  triple points for small and heavy quark masses,
and second order points for intermediate quark masses. So there exist two tricritical points  separating the first order
transition points from the second ones. Moreover, it is pointed out~\cite{deForcrand:2010he,Philipsen:2010rq,Bonati:2012pe} that the scaling behaviour at the tricritical points
may shape the  critical line which separate different transition region for real chemical potential, and thus, the critical line for real chemical potential is expected to be
qualitatively consistent with the scenario suggested in Refs.~\cite{deForcrand:2006pv,deForcrand:2008vr} which shows that the first order transition region shrinks with increasing real chemical potential. In addition, Ref.~\cite{Bonati:2014kpa} employs the scaling behaviour at the tricritical point to determine the nature of 2 flavour QCD transition in the chiral limit.

So far, the investigation for the Roberge-Weiss transition endpoints are implemented through standard gauge and fermion actions.
In this paper, we aim to investigate the endpoints  of $N_f=2$ QCD with one-loop Symanzik-improved gauge action
\cite{Symanzik:1983dc,Luscher:1985zq,Lepage:1992xa,Alford:1995hw}
and Asqtad  KS action \cite{Blum:1996uf,Orginos:1999cr}.  These actions have discretization error of $O(\alpha_s^2 a^2, a^4)$ and $O(\alpha_s a^2, a^4)$, respectively.
These improvements are significant on $N_t=4$ lattice where the lattice spacing is quite large.    Standard KS fermions suffer from taste symmetry breaking at nonzero lattice spacing $a$ \cite{Bazavov:2011nk}. This taste symmetry breaking can be illustrated by the smallest pion mass taste splitting which is comparable to the pion mass even at lattice spacing $a \sim 0.1fm$ \cite{Bazavov:2009bb}.
Asqtad KS action has good taste symmetry and free dispersion relation by introducing
fattened links and the so-called "Naik terms" \cite{Naik:1986bn,Bernard:1997mz}.

The paper is organized as follows. In Sec.~\ref{SectionLattice},
we define the lattice action with imaginary chemical potential and
the physical observables we calculate.  Our simulation results are
presented in Sec.~\ref{SectionMC} followed by discussions in
Sec.~\ref{SectionDiscussion}.

\begin{figure*}[t!]
\includegraphics*[width=0.49\textwidth]{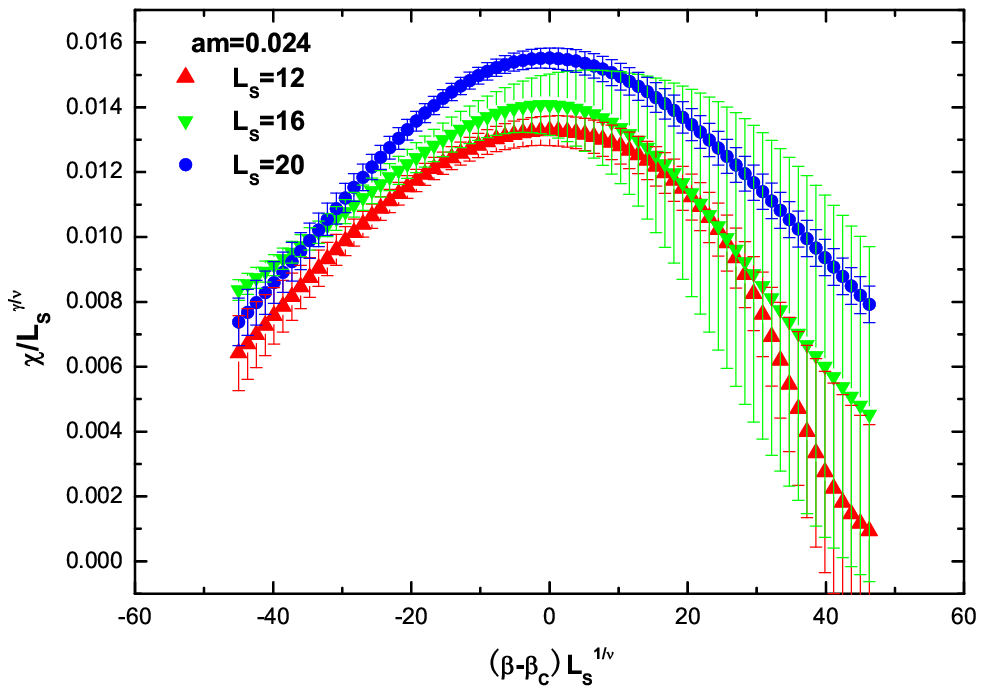}
\includegraphics*[width=0.49\textwidth]{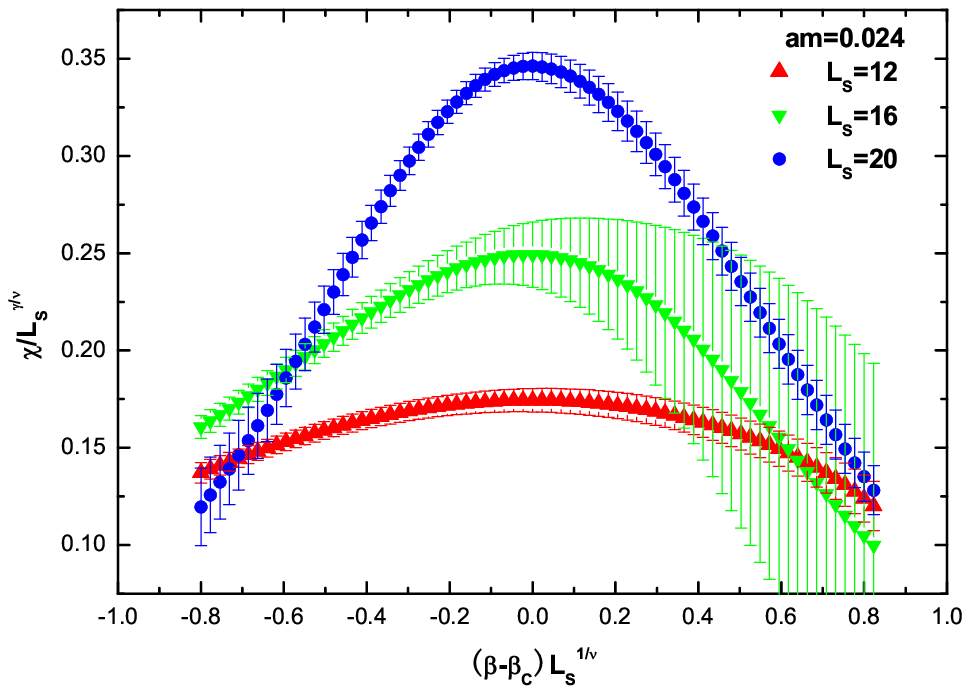}\\
\caption{\label{fig1} Scaling behavior of the susceptibility of  imaginary part of  Polyakov loop
according to  first order critical index (left panel), and to  3D Ising critical index (right panel) at $am=0.024$.}
\end{figure*}

\begin{figure*}[t!]
\includegraphics*[width=0.49\textwidth]{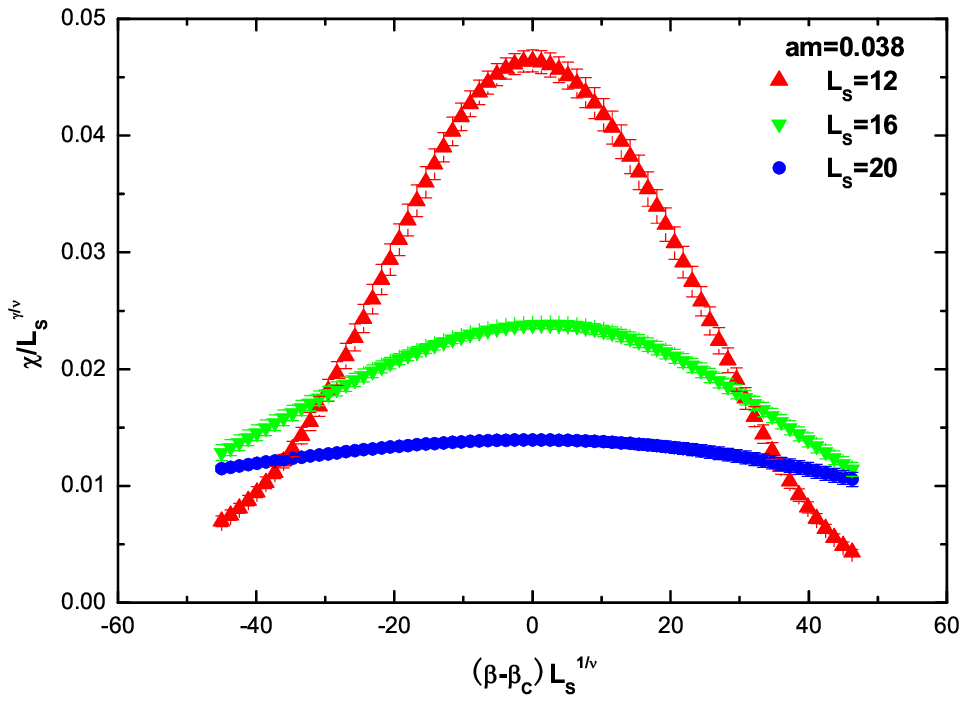}
\includegraphics*[width=0.49\textwidth]{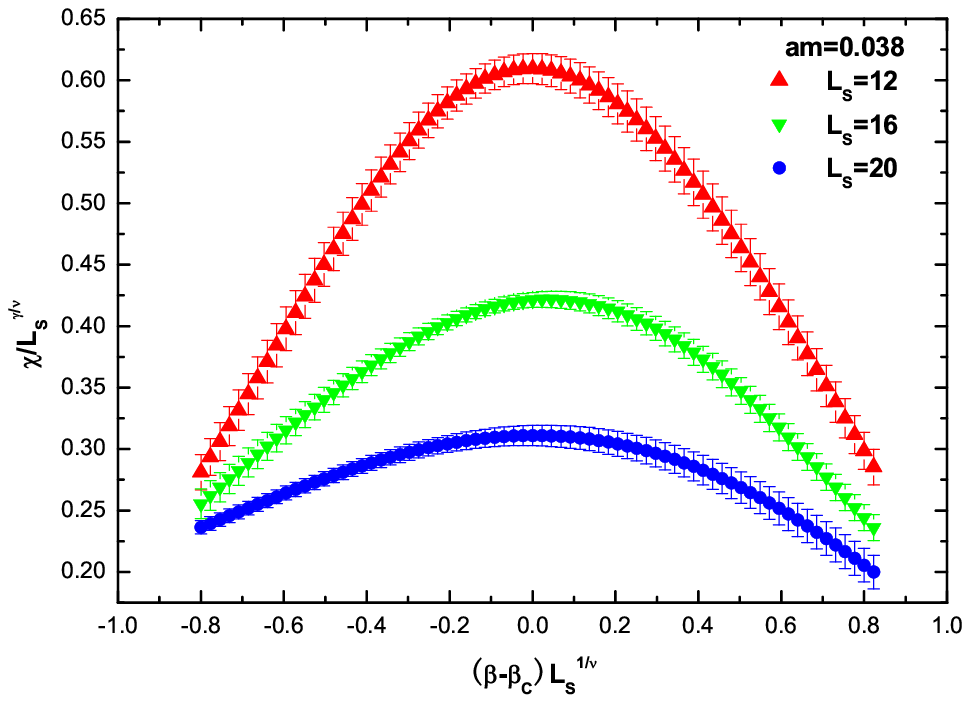}
\caption{\label{fig2} Scaling behavior of the susceptibility of imaginary part of  Polyakov loop
according to  first order critical index (left panel), and to  3D Ising critical index (right panel) at $am=0.038$.}
\end{figure*}

\begin{figure*}[t!]
\includegraphics*[width=0.49\textwidth]{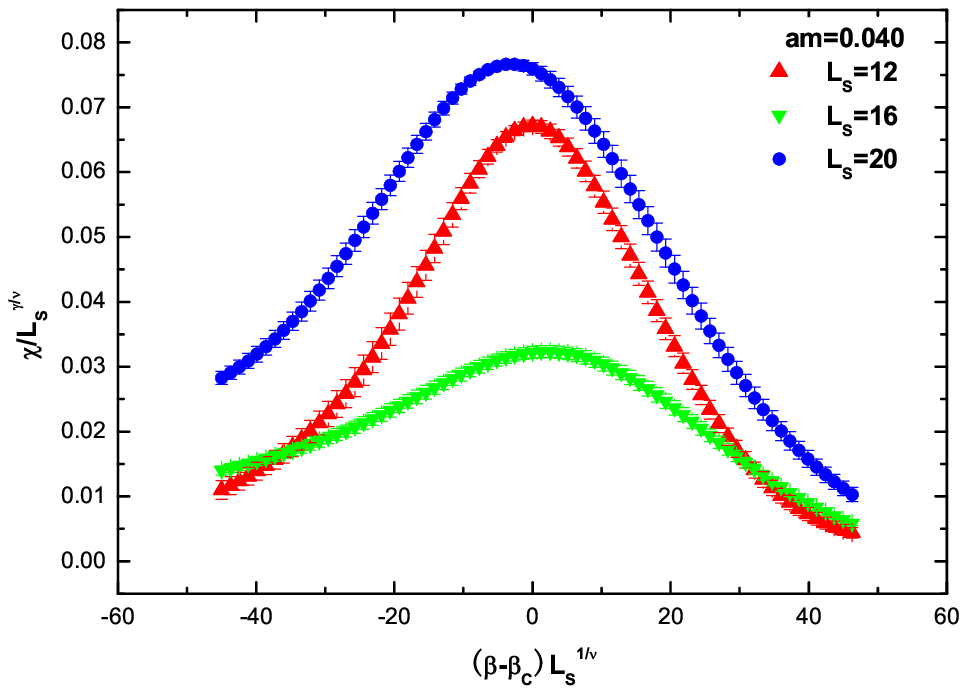}
\includegraphics*[width=0.49\textwidth]{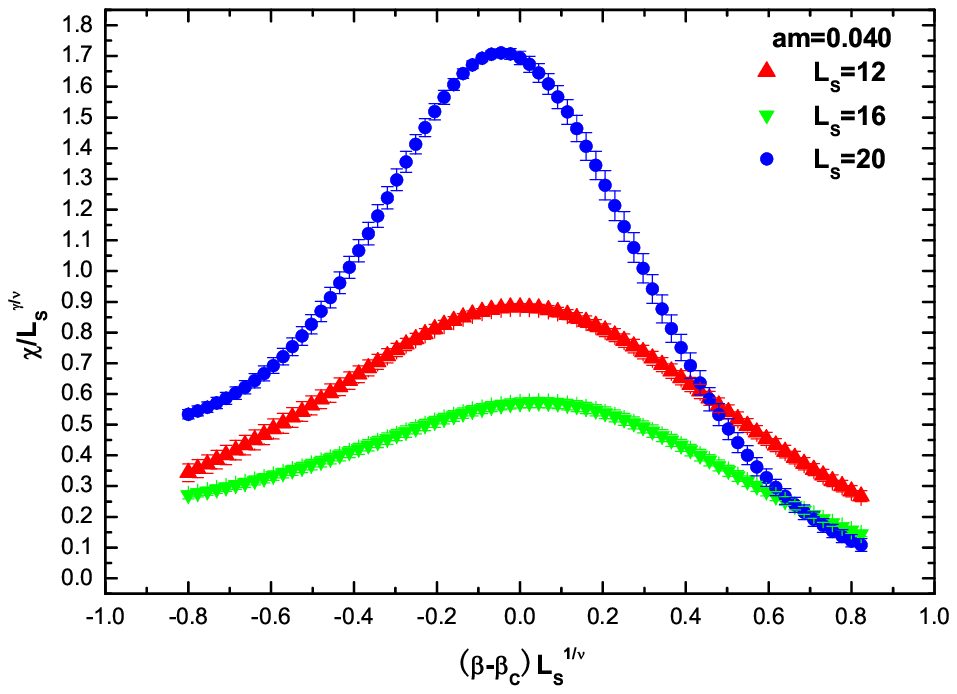}
\caption{\label{fig3} Scaling behavior of the susceptibility of imaginary part of  Polyakov loop
according to  first order critical index (left panel), and to  3D Ising critical index (right panel) at $am=0.040$.}
\end{figure*}

\begin{figure*}[t!]
\includegraphics*[width=0.49\textwidth]{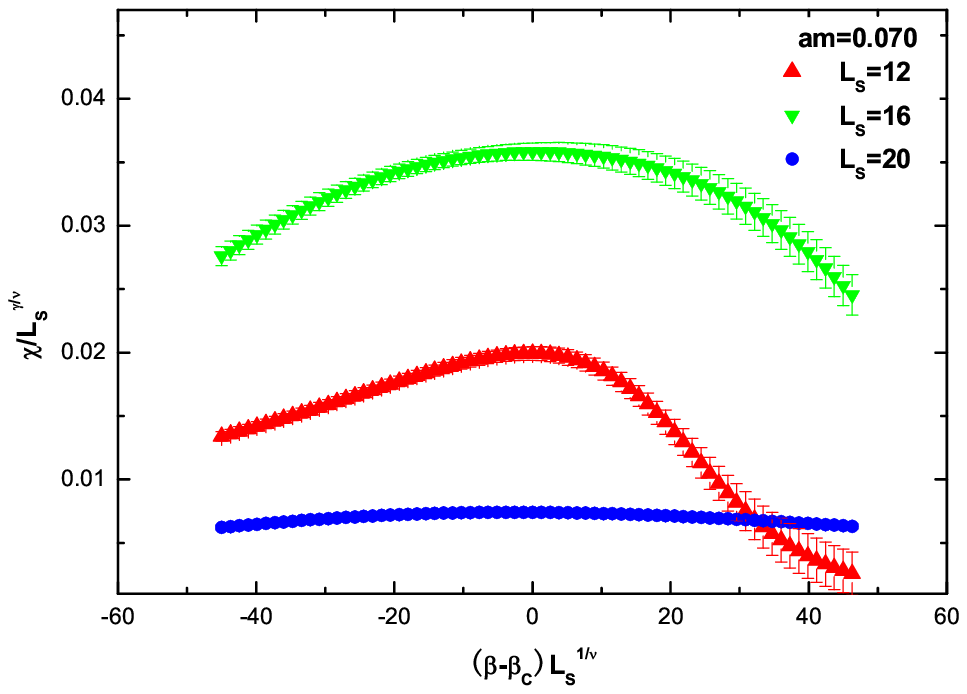}
\includegraphics*[width=0.49\textwidth]{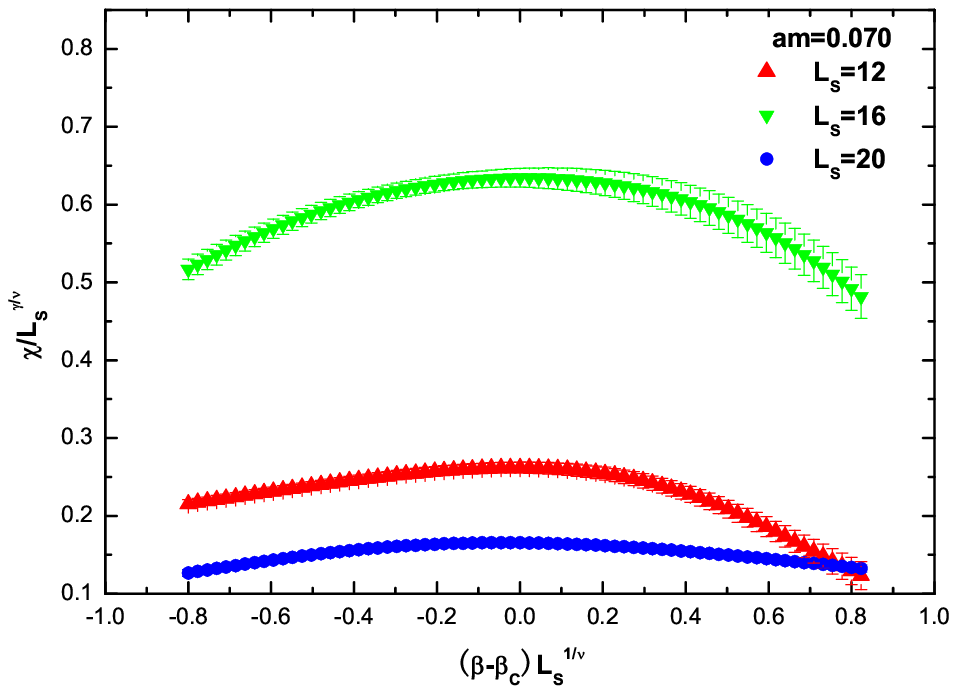}
\caption{\label{fig4} Scaling behavior of the susceptibility of  imaginary part of  Polyakov loop
according to  first order critical index (left panel), and to  3D Ising critical index (right panel) at $\kappa=0.070$.}
\end{figure*}

\begin{figure*}[t!]
\includegraphics*[width=0.49\textwidth]{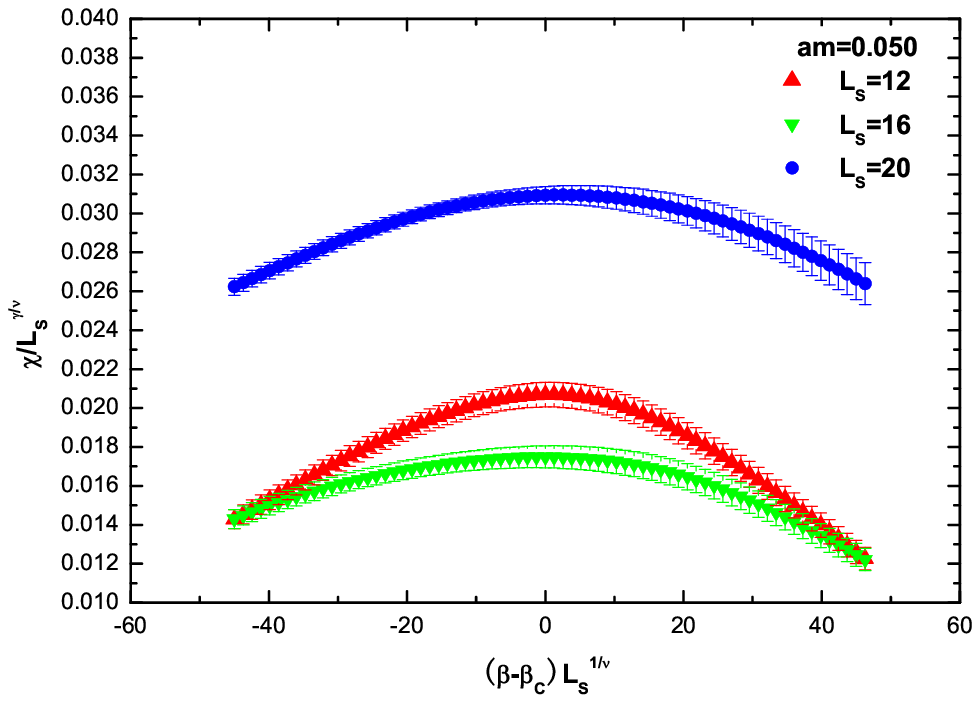}
\includegraphics*[width=0.49\textwidth]{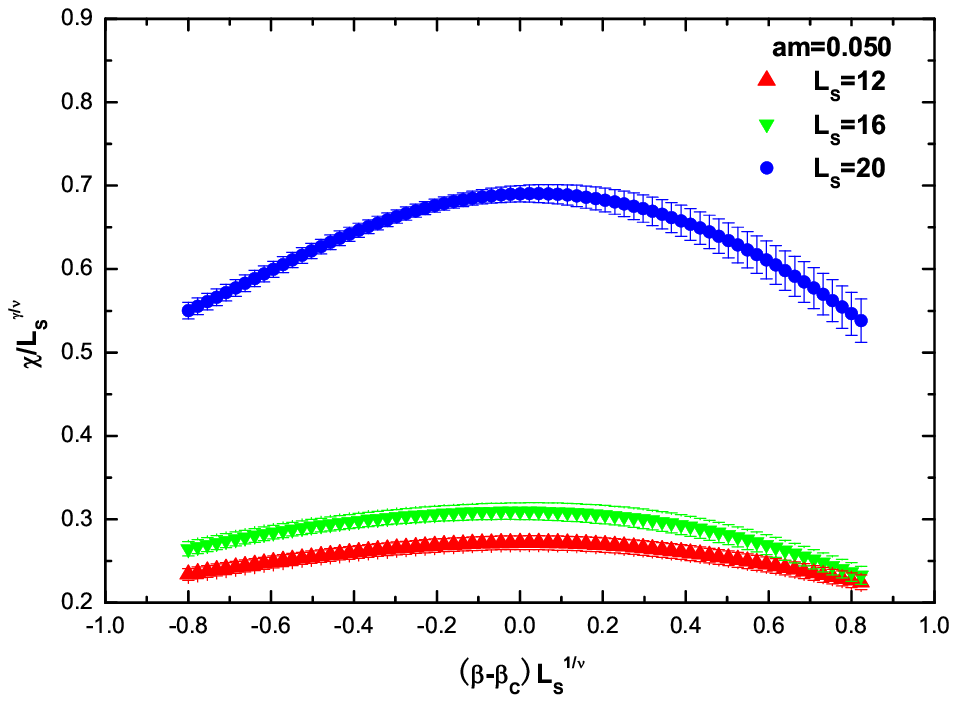}
\caption{\label{fig5} Scaling behavior of the susceptibility of  imaginary part of  Polyakov loop
according to  first order critical index (left panel), and to 3D Ising critical index (right panel) at $\kappa=0.050$.}
\end{figure*}

\begin{figure*}[t!]
\includegraphics*[width=0.49\textwidth]{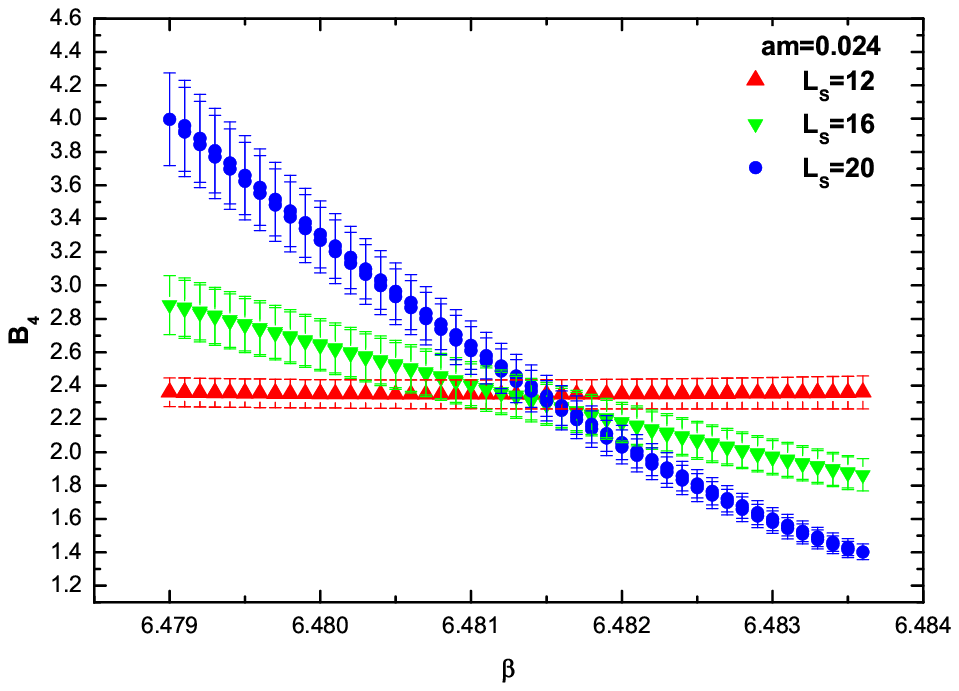}
\includegraphics*[width=0.49\textwidth]{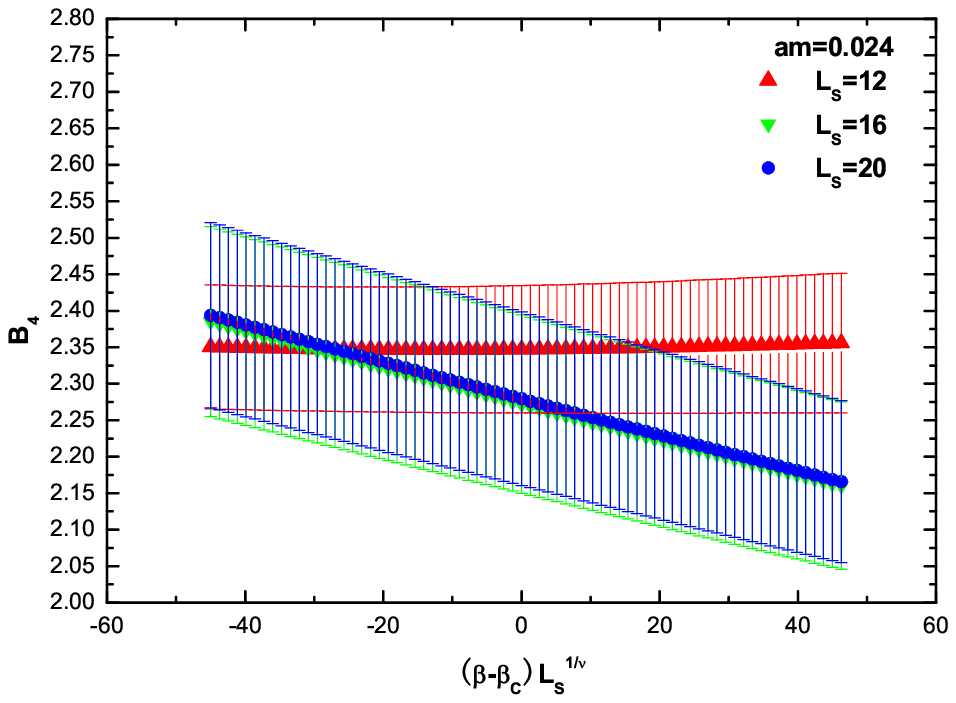}
\caption{Binder cumulants as  a function of $\beta$ on various spatial volume intersect at one point (left panel),
and as  a function of $(\beta-\beta_c)L_s^{1/\nu}$ with values of $\beta_c$, $\nu$ from Table.~\ref{critical_beta_B4} collapse (right panel) at $am=0.024$.}
\label{fig6}
\end{figure*}

\begin{figure*}[t!]
\includegraphics*[width=0.49\textwidth]{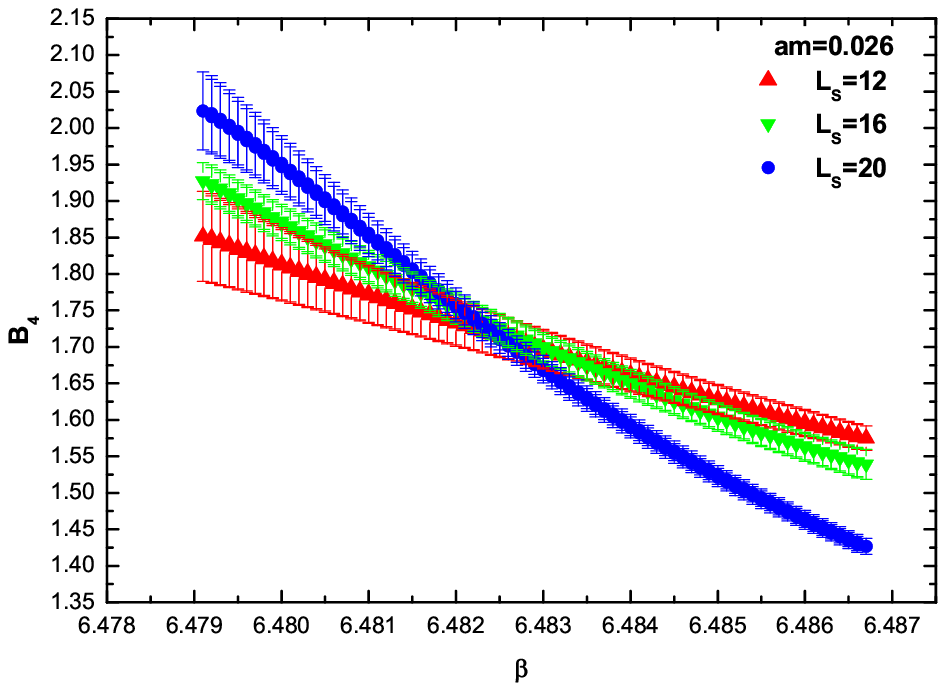}
\includegraphics*[width=0.49\textwidth]{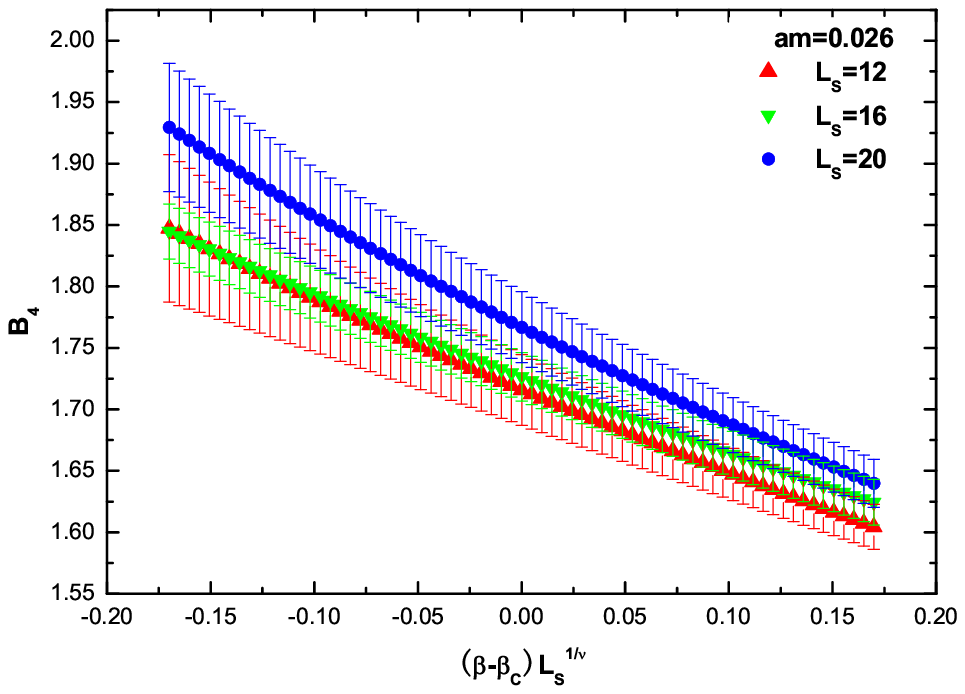}
\caption{\label{fig7} Binder cumulants as  a function of $\beta$ on various spatial volume intersect at one point (left panel),
and as  a function of $(\beta-\beta_c)L_s^{1/\nu}$ with values of $\beta_c$, $\nu$ from Table.~\ref{critical_beta_B4} collapse (right panel) at $am=0.026$.}
\end{figure*}

\begin{figure*}[t!]
\includegraphics*[width=0.49\textwidth]{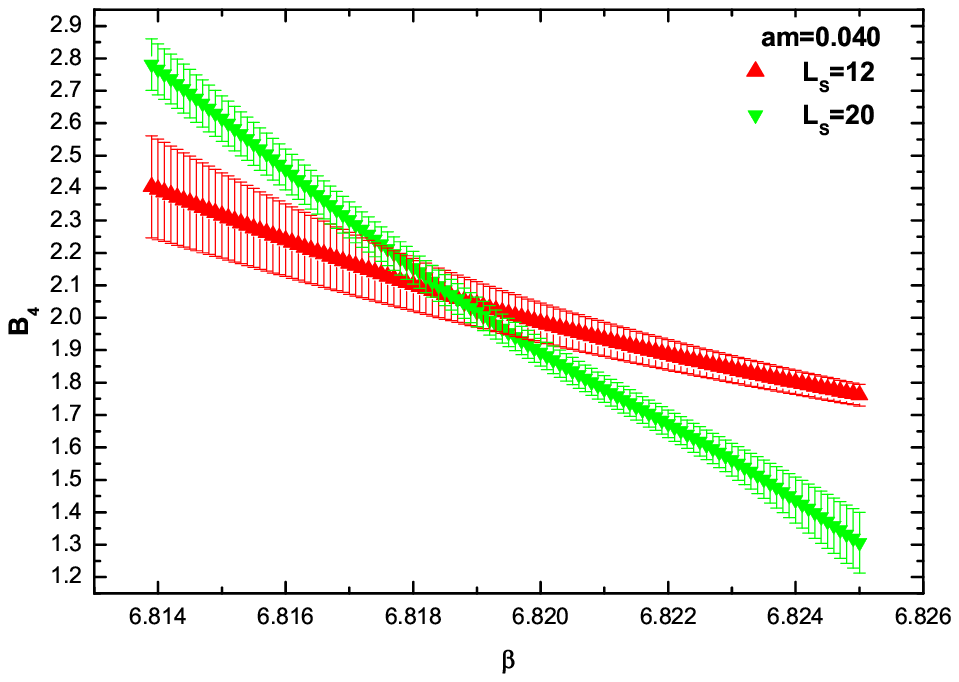}
\includegraphics*[width=0.49\textwidth]{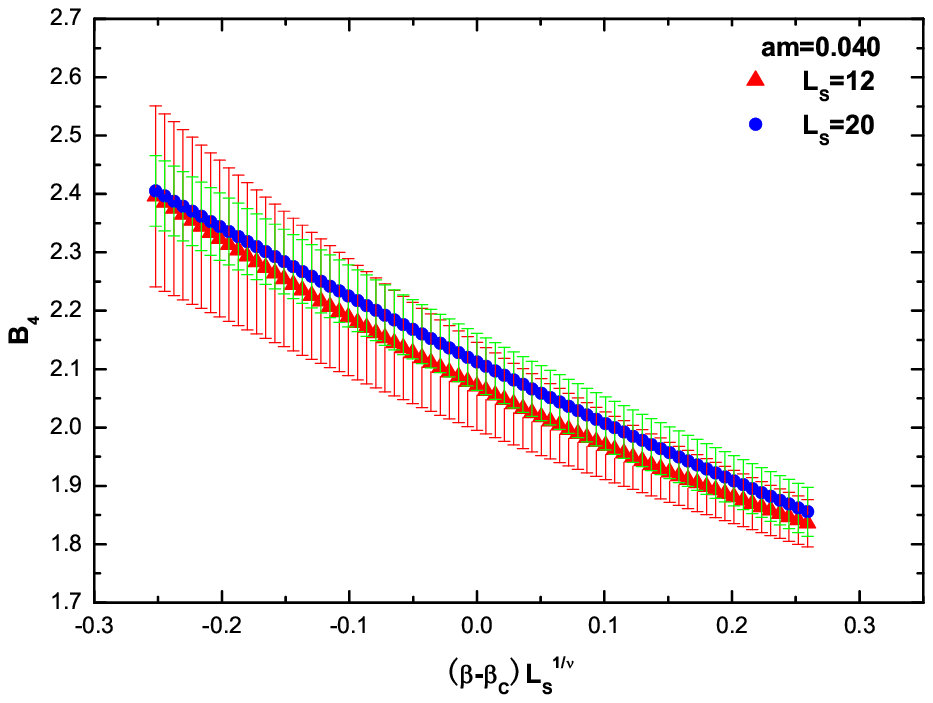}
\caption{\label{fig8} Binder cumulants as  a function of $\beta$ on various spatial volume intersect at one point (left panel),
and as  a function of $(\beta-\beta_c)L_s^{1/\nu}$ with values of $\beta_c$, $\nu$ from Table.~\ref{critical_beta_B4} collapse (right panel) at $am=0.040$. }
\end{figure*}

\section{LATTICE FORMULATION WITH IMAGINARY CHEMICAL POTENTIAL}
\label{SectionLattice} After introducing pseudofermion field $\Phi$, the partition function of the system can be represented as:
       \begin{eqnarray}
        \label{QCD_partition}
          Z &= &\int [dU][d\Phi^*][d\Phi]e^{-S_g-S_f}, \nonumber
       \end{eqnarray}
where $S_g$ is the Symanzik-improved gauge action, and $S_f$ is the Asqtad quark
action with the quark chemical potential $\mu$. Here $\mu=
i\mu_I$. For  $S_g$, we use

\newlength{\latlength}
\setlength{\latlength}{1mm}
\setlength{\unitlength}{\latlength}
\ba
  S_G = \beta \left( C_P \sum_{x;\mu<\nu} (1 -  P_{\mu\nu})
      + C_R \sum_{x;\mu\neq\nu} (1 - R_{\mu\nu}) \right.  \nonumber \\
 \left.      + C_T \sum_{x;\mu<\nu<\sigma} (1 -  T_{\mu\nu\sigma}) \right), \nonumber
 \label{gaugeaction}
\ea
with  $P_{\mu\nu}, R_{\mu\nu}$ and $ T_{\mu\nu\sigma}$ standing for $1/3$ of the imaginary part of the trace of $1\times1$, $1\times2$ planar Wilson loops and  $1\times1\times1$ "parallelogram" loops, respectivley.
\ba
   P_{\mu\nu} &=& \frac{1}{3} {\rm Re \rm Tr}\
	\raisebox{-4\latlength}{\begin{picture}(10,12)
			\put( 0, 0){\vector( 1, 0){7}}
			\put( 0, 0){\line( 1, 0){10}}
			\put(10, 0){\vector( 0, 1){7}}
			\put(10, 0){\line( 0, 1){10}}
			\put(10,10){\vector(-1, 0){7}}
			\put(10,10){\line(-1, 0){10}}
			\put( 0,10){\vector( 0,-1){7}}
			\put( 0,10){\line( 0,-1){10}}
		\end{picture}}\ ,  \nonumber
\ea
\ba
   R_{\mu\nu} &=& \frac{1}{3} {\rm Re \rm Tr}\
	\raisebox{-4\latlength}{\begin{picture}(20,12)
			\put( 0, 0){\vector( 1, 0){7}}
			\put( 0, 0){\line( 1, 0){10}}
			\put(10, 0){\vector( 1, 0){7}}
			\put(10, 0){\line( 1, 0){10}}
			\put(20, 0){\vector( 0, 1){7}}
			\put(20, 0){\line( 0, 1){10}}
			\put(20,10){\vector(-1, 0){7}}
			\put(20,10){\line(-1, 0){10}}
			\put(10,10){\vector(-1, 0){7}}
			\put(10,10){\line(-1, 0){10}}
			\put( 0,10){\vector( 0,-1){7}}
			\put( 0,10){\line( 0,-1){10}}
			\multiput(10, 1)(0,1){9}{\circle*{0.1}}
		\end{picture}}\ ,  \nonumber
\ea
\ba
   T_{\mu\nu\sigma} &=& \frac{1}{3} {\rm Re \rm Tr}\
	\raisebox{-4\latlength}{\begin{picture}(16,16)
			\put( 0, 0){\vector( 1, 0){7}}
			\put( 0, 0){\line( 1, 0){10}}
			\put(10, 0){\vector( 2, 1){4}}
			\put(10, 0){\line( 2, 1){6}}
			\put(16, 3){\vector( 0, 1){7}}
			\put(16, 3){\line( 0, 1){10}}
			\put(16,13){\vector(-1, 0){7}}
			\put(16,13){\line(-1, 0){10}}
			\put( 6,13){\vector(-2,-1){4}}
			\put( 6,13){\line(-2,-1){6}}
			\put( 0,10){\vector( 0,-1){7}}
			\put( 0,10){\line( 0,-1){10}}
			\multiput(1,0.5)(1,0.5){6}{\circle*{0.1}}
			\multiput(6,4)(0,1){9}{\circle*{0.1}}
			\multiput(7,3)(1,0){9}{\circle*{0.1}} \nonumber
		\end{picture}}\ .
\ea

The coefficents $C_P ,C_R, C_T $  are tadpole improved~\cite{Bazavov:2009bb},
\ba
C_P &=& 1.0, \nonumber \\
C_R &=& \frac{-1}{20 u_0^2} \left( 1 - \left( 0.6264 - 1.1746 n_f \right) {\rm ln}(u_0) \right),  \nonumber \\
C_T &=& \frac{1}{u_0^2} \left( 0.0433 - 0.0156 n_f \right) {\rm ln}(u_0).  \nonumber
\ea

The Asqtad action with pseudofermion field $\Phi$ is
\begin{equation}
S_{f} = \left< \Phi\left|\left[
M^{\dag}[U]M[U]
\right]^{-n_{f}/4} \right|\Phi \right>, \nonumber
\end{equation}
where the form of $M_{x,y}\left[ U \right]=2m_{x,y}+D_{x,y}(U)$  reading 
\begin{eqnarray}
&2m\delta_{x,y}+\sum\limits_{\rho=1}^{3}\eta_{x,\rho}
\left( U_{x,\rho}^{F}\delta_{x,y-\hat\rho}-U^{F\dag}_{x-\hat\rho,\rho}\delta_{x,y+\hat\rho}
\right)       \nonumber \\
+&\eta_{x,4}\left( e^{ia\mu_I}U_{x,4}^{F}\delta_{x,y-\hat4}-e^{-ia\mu_I} U^{F\dag}_{x-\hat4,\mu}\delta_{x,y+\hat4}
\right) \nonumber \\
+&\sum\limits_{\rho=1}^{3}\eta_{x,\rho}
\left( U_{x,\rho}^{L}\delta_{x,y-3\hat\rho}-U^{L\dag}_{x-\rho,\rho}\delta_{x,y+3\hat\rho}
\right) \nonumber \\
+&\eta_{x,4}\left( e^{i3a\mu_I}U_{x,4}^{L}\delta_{x,y-3\hat4}-e^{-i3a\mu_I} U^{L\dag}_{x-\hat4,\mu}\delta_{x,y+3\hat4}
\right), \nonumber
\end{eqnarray}
where $U_{x,\rho}^{F}$ stands for the fattened link which is produced by Fat7 smearing and $U_{x,\rho}^{L}$ stands for the naik term.
$\hat\rho, \,\hat 4$ are the unit vector along $\rho-$direction,$4-$direction, respectively. $\eta_{x,\mu}$ is the staggered fermion phase.

We carry out simulations at $\theta=\mu_I/T=\pi$. As it is pointed out that the system is invariant under the charge
 conjugation at $\theta =0,\pi$, when $\theta $ is fixed~\cite{Kouno:2009bm}. But the $\theta$-odd quantity $O(\theta)$ is not invariant at $\theta=\pi$
 under charge conjugation. When $T < T_{RW}$, $O(\theta)$ is a smooth function of $\theta$, so it is zero at $\theta=\pi$.
 Whereas when $T > T_{RW}$,  the two charge violating solutions cross each other at $\theta=\pi$. Thus the charge symmetry is
 spontaneously broken there and the $\theta$-odd quantity $O(\theta)$ can be taken as order parameter . In this paper, we take the imaginary part of Polyakov loop as the order parameter.

The expression of  Polyakov loop $ L $ is defined as the following:
\begin{eqnarray}
\langle  L \rangle=\left\langle \frac{1}{3L_s^3L_t}\sum_{\bf x}{\rm  Tr}
\left[ \prod_{t=1}^{L_t} U_4({\bf x},t) \right] \right\rangle ,\nonumber
\end{eqnarray}
$L_s, L_t$ are the spatial, time extent of lattice, respectively. To simplify notation, we use $X$ to represent
the imaginary part of Polyakov loop $ {\rm Im}( L )$.
The susceptibility of imaginary part of Polyakov loop $ {\rm Im}( L )$   is defined as
\begin{eqnarray}
\chi= L_s^3\left\langle( X- \langle  X\rangle)^2\right\rangle, \nonumber
\end{eqnarray}
which is expected to scale as:~\cite{D'Elia:2009qz,Bonati:2010gi}
\begin{eqnarray}\label{chi_scaling}
\chi= L_s^{\gamma/\nu}\phi(\tau L_s^{1/\nu}),
\end{eqnarray}
where $\tau $ is the reduced temperature $\tau=(T-T_{RW})/T_{RW}$.
This means that the curves $\chi/L_s^{\gamma/\nu}$ at different lattice volume should collapse with the same curve when plotted against $\tau L_s ^{1/\nu}$.
In the following, we employ $\beta-\beta_{RW}$ in place of $\tau=(T-T_{RW})/T_{RW}$. The critical exponents relevant to our study
are collected in Table.~\ref{critical_exponents}~\cite{Bonati:2010gi,Pelissetto:2000ek}.
\begin{table}[htp]
\begin{ruledtabular}
\begin{center}
\begin{tabular}{ccccc}
              &$B_4(\beta_c,\infty)$    &  $\nu$     & $\gamma $    &       $\gamma/\nu$    \\
 \hline
3D ising      &1.604                    & 0.6301(4)  & 1.2372(5)    &        1.963                  \\
tricritical   &2                        & 1/2        &  1           &           2                         \\
 first order  & 1.5                     & 1/3        &  1           &               3                        \\
 crossover    & 3                       & -          &  -           &              -                        \\
\end{tabular}
\end{center}
\end{ruledtabular}
\caption{\label{critical_exponents}Critical exponents relevant to our study.}
\end{table}

We also consider the Binder cumulant of  imaginary part of Polyakov loop which is defined as the following:
\begin{eqnarray}\label{binder_scaling}
B_4=\left\langle ( X - \langle  X\rangle)^4\right\rangle /
    \left\langle ( X - \langle  X\rangle)^2\right\rangle^2
\end{eqnarray}
with  $\langle X \rangle =0$.
In the vicinity of the RW transition line
endpoints, $B_4$ with the finite size correction is a function of $x=(\beta-\beta_{RW})L_s^{1/\nu}$ and can be expanded as a series~\cite{deForcrand:2010he,Philipsen:2010rq,Bonati:2012pe},
\begin{eqnarray}\label{binder_scaling_02}
B_4=B_4(\beta_c,\infty)+a_1x+a_2x^2+\cdots.
\end{eqnarray}
 In the thermodynamic limit, the critical index $\nu$ and $B_4(\beta_c,\infty)$ takes on the corresponding value summarized in Table.~\ref{critical_exponents}.
However, on finite spatial volumes, the steps of $B_4(\beta_c,\infty)$ are smeared out to continuous functions.

\section{MC SIMULATION RESULTS}
\label{SectionMC}

Before presenting the simulation results, we describe the simulation details.    Simulations are carried out at quark mass $am=0.024,\,$ $0.026,\,$  $0.038,\,$  $0.040,\,$  $0.050,\,$  $0.060,\,$  $0.070$.   Rational Monte Carlo algorithm \cite{Clark:2003na,Clark:2006wp,Clark:2006fx}
is used to generate configurations. The Omelyan integration algorithm \cite{Takaishi:2005tz,omeylan}
is employed for the gauge and fermion action. For the molecular dynamics evolution we use a 9'th rational function to approximate $[M^+(U)M(U)]^{-n_f/4}$ for the pseudofermion field. For the heat bath updating and for computing the action at the beginning and end of the molecular dynamics trajectory 10'th rational function is  used to approximate $[M^+(U)M(U)]^{n_f/8}$ and $[M^+(U)M(U)]^{-n_f/8}$, respectively. The step is chosen to ensure the acceptance rate is around $80\%-90\% $. 5,000 trajectories of configuration are taken as warmup form a cold start. 
In order to fill in  observables   at additional $\beta$ values,
we employ the Ferrenberg-Swendsen reweighting
method~\cite{Ferrenberg:1989ui}.

The critical coupling $\beta_{RW}$'s on various spatial volume at different quark mass $am$ are summarized in Table.~\ref{critical_beta}. These $\beta_{RW}$'s are determined from the locations of  peak susceptibility of imaginary part of Polyakov loop.
\begin{table}[htp]
\caption{\label{critical_beta}Results of critical couplings $\beta_{RW}$ on different spatial volume at different $\kappa$. }
\begin{ruledtabular}
\begin{center}
\begin{tabular}{c|cccc}

$am$        &     $12 $        &         $  16  $           &    $20 $             \\  \hline
0.024       &   $6.492(9) $    &          $6.491(8)$    & $6.4834(15)$     \\
0.038       & $6.838(4) $      &         $ 6.821(4) $       & $6.824(3) $                 \\
0.040       & $ 6.839(3)  $     &           $ 6.839(3)$      &  $6.847(2)$      \\
0.050       & $6.845(10) $     &            $6.831(7)$      & $ 6.857(4)$     \\
0.060       & $6.859(9)$       &           $6.865(14)$      & $ 6.860(3)$      \\
0.070       & $6.875(7)$       &            $6.885(6)$      & $ 6.857(4)$      \\
\end{tabular}
\end{center}
\end{ruledtabular}
\end{table}


\begin{table*}[htp]
\caption{\label{critical_beta_B4}Results of critical couplings $\beta_{RW}$ and the critical index $\nu$ by fitting Eq.~(\ref{binder_scaling_02}) to data on different spatial volume. If errors  are very small, we take them to be zero.   }
\begin{ruledtabular}
\begin{center}
\begin{tabular}{c|ccccccc}

$am$    &     $L_s $      &   $  \beta_{RW}$  & $\nu $         & $  B_4(\beta_c,\infty) $  & $a_1$  &  $a_2$  & r-square    \\  \hline
0.024       &$12 \,,16 \,,20$        &   $ 6.4816(0)$    & $0.2410(8)$    & $ 2.2661(11)$   &  $-0.0022(0)$ & $0.000(0)$  &   0.991  \\
0.026      &$12 \,,16\,, 20$        &   $ 6.4825(0)$    & $0.6282(3)$    & $ 1.71958(6)$   &  $-0.7061(14)$  & $0.2033(9)$    & 0.996 \\
0.038       &$16\,, 20$        &   $ 6.8503(0) $    & $0.6473(17)$    & $ 1.0300(0)$   &  $-0.0363(4)$ & $0.01145(2)$     &   0.996  \\
0.040       &$12,\, 20$        &   $ 6.8185(0)$    & $ 0.6173(4)$   & $2.1039(3)$    &  $-1.053(3)$ & $0.136(8)$ &   0.998  \\
0.050       &$12,\,16,\, 20$        &   $ 6.831(0)$    & $0.3691(6)$    & $ 1.8924(2)$   &  $-0.0295(4)$ & $0.0008(0)$     &   0.992  \\
0.060       &$12,\, 20$        &   $6.8416(0)$    & $0.3458(19)$    & $ 1.6937(10)$   &  $-0.0125(6)$ & $-$     &   0.958  \\
0.070       &$12\,, 20\     $        &   $ 6.8416(0)$   & $ 0.3152(6)$   & $2.1821(2)$    &  $-0.005(0)$ & $-$ &   0.936 \\
\end{tabular}
\end{center}
\end{ruledtabular}
\end{table*}

 We present the rescaling susceptibility of imaginary part of Polyakov loop $\chi/{{L_s}^{\gamma/\nu}}$  as a function of $(\beta-\beta_{RW})L_s^{1/\nu}$ at $am=0.024$ in Fig.~\ref{fig1}. From  Fig.~\ref{fig1}, we can find that
$\chi/{{L_s}^{\gamma/\nu}}$ according to the first order transition index  collapses with the same curve, while
$\chi/{{L_s}^{\gamma/\nu}}$ according to 3D index  does not.

The rescaling susceptibility of imaginary part of Polyakov loop $\chi/{{L_s}^{\gamma/\nu}}$  as a function of $(\beta-\beta_{RW})L_s^{1/\nu}$ at $am=0.038$  is depicted in Fig.~\ref{fig2}. From  Fig.~\ref{fig2}, we can find that
$\chi/{{L_s}^{\gamma/\nu}}$ according to  the first order transition index  or 3D index  does not collapse with the same curve.  We cannot determine the nature of Roberge-Weiss transition endpoint at $am=0.038$ from $\chi/{{L_s}^{\gamma/\nu}}$.

The behaviour of rescaling susceptibility of imaginary part of Polyakov loop $\chi/{{L_s}^{\gamma/\nu}}$ at $am=0.040$  and $am=0.070$ are presented in Fig.~\ref{fig3}, and Fig.~\ref{fig4} respectively. Form Fig.~\ref{fig3} and Fig.~\ref{fig4}, we can find that  The rescaling susceptibility of imaginary part of Polyakov loop $\chi/{{L_s}^{\gamma/\nu}}$ at $am=0.040$  and $am=0.070$ have similar behaviour to the that at $am=0.038$.

The rescaling susceptibility of imaginary part of Polyakov loop $\chi/{{L_s}^{\gamma/\nu}}$ as a function of $(\beta-\beta_{RW})L_s^{1/\nu}$ at $am=0.050$ is depicted in Fig.~\ref{fig5}.  From Fig.~\ref{fig5}, we can find that  $\chi/{{L_s}^{\gamma/\nu}}$ as a function of $(\beta-\beta_{RW})L_s^{1/\nu}$ at lattice $12^3\times4$ and $16^3\times 4$ are in favour of both first order transition index and 3D index. However, considering the the scale of $\chi/{{L_s}^{\gamma/\nu}}$ and $(\beta-\beta_{RW})L_s^{1/\nu}$ in
Fig.~\ref{fig5}, the first order transition index may be the better choice.  $\chi/{{L_s}^{\gamma/\nu}}$ as a function of $(\beta-\beta_{RW})L_s^{1/\nu}$ at $am=0.060$ has similar behaviour to that at $am=0.050$ which tends to be  in favour of  first order transition index.

 In order to discern the scaling behaviour,
we turn to investigate Binder cumulant $B_4$ as defined in Eq.~(\ref{binder_scaling}) whose scaling behaviour is described in Eq.~(\ref{binder_scaling_02}). $B_4$ decreases with the increase of   $\beta$, and at one fixed quark mass $am$, $B_4$ as a function of $\beta$ on various spatial volume is expected to intersect at one point. The intersection gives an estimate of accurate location of $\beta_{RW}$. By  fitting  to
Eq.~(\ref{binder_scaling_02}), we can extract critical index $\nu, \, \beta_{RW}$ and $B_4(\beta_c,\infty)$. The results are collected in Table.~\ref{critical_beta_B4}.

We present $B_4$ as a function of $\beta$  at $am=0.024$ in the left panel of Fig.~\ref{fig6},
and  $B_4$ as a function of $(\beta-\beta_{RW})L_s^{1/\nu}$ in the right panel of Fig.~\ref{fig6} with $\nu$ taken to be the extracted value through  fitting procedure. From Table.~\ref{critical_beta_B4},  we find that the critical index $\nu=0.2410$ at $am=0.024$ can explain the behaviour of $B_4$ as a function of $(\beta-\beta_{RW})L_s^{1/\nu}$, especially, on lattice $L_s=16,20$. This behaviour implies that the transition  endpoint at $am=0.024$ belongs to first order transition.

We also present $B_4$ as a function of $\beta$  at $am=0.026$ in the left panel of Fig.~\ref{fig7},
and  $B_4$ as a function of $(\beta-\beta_{RW})L_s^{1/\nu}$ in the right panel of Fig.~\ref{fig7} with $\nu$ taken to be the extracted value through  fitting procedure. We find that the critical index $\nu=0.6282$ at $am=0.026$ can explain the behaviour of $B_4$ as a function of $(\beta-\beta_{RW})L_s^{1/\nu}$. $\nu=0.6282$ suggests that the endpoint at $am=0.026$ is of 3D transition nature.

At $am=0.040$, we only find that $B_4$ as a function of $\beta$ on lattice $L_s=12,20$ intersects at one point. $B_4$ as a function of $\beta$  and as a function of $(\beta-\beta_{RW})L_s^{1/\nu}$ at $am=0.040$ are depicted in the left, right panel of Fig.~\ref{fig7}, respectively. The extracted value $\nu=0.6173$ through  fitting procedure also shows that the endpoint at $am=0.040$ is of 3D transition nature. At other values of $am$, $B_4$ as a function of $\beta$ and  as a function of $(\beta-\beta_{RW})L_s^{1/\nu}$ have similar behaviour. For clarity, they are not presented.

From the behaviour of $\chi/{{L_s}^{\gamma/\nu}}$  and $B_4$,   we conclude that the nature of endpoint transition at $am=0.024,$ $0.050,$
$0.060,$ $0.070$ is of first order, while at $am=0.026,$ $0.038,$
$0.040,$ the endpoint transition nature is of 3D Ising class. This conclusion suggests that the two tricritical points are between $0.024<am_{tricl}<0.026$ and $0.040<am_{tricl}<0.050$.

\section{DISCUSSIONS}\label{SectionDiscussion}

We have studied the nature of critical endpoints of Roberge-Weiss transition of two flavor lattice QCD with
improved KS fermions. When $i\mu_I = i\pi T$, the imaginary part of Polayakov loop is the order parameter for studying
the transition from low temperature phase to high temperature one.

 Our simulations are carried out at 7  values of quark mass $am$ on $L_t=4 $ lattice on different 3 spatial volumes. Our central result is that the two tricritical points are between $0.024<am_{tricl}<0.026$ and $0.040<am_{tricl}<0.050$. The interval of quark mass from
 0.024 to 0.026 is narrow. On finite spatial volume, the index $\nu$ is expected to change smoothly, while our simulation shows that index $\nu$ changes rapidly within a narrow quark mass interval.

 In Ref.~\cite{Bonati:2010gi}, the two locations of tricritical point for $N_f=2$ QCD are $am=0.043(5),\,0.72(8)$, respectively.
 For $N_f=3$ QCD, Ref.~\cite{deForcrand:2010he} concludes that the two tricritical points are between $0.07<am_{tricl}<0.3$ and $0.5<am_{tricl}<1.5$. Comparing with those results, the second transition region from our simulation is narrow.

 Apart from monitoring the behaviour of susceptility of imaginary part of Polayakov loop ${\rm Im}(L)$, we also look into the change of Binder cumulant of ${\rm Im}(L)$. 
 In order to fill in  observables   at additional $\beta$ values, the Ferrenberg-Swendsen reweighting
method~\cite{Ferrenberg:1989ui} is employed. It is noted that when applying Ferrenberg-Swendsen reweighting method, the number of $\beta$ points taken to calculate susceptility is not completely the same as the number taken to calculate Binder cumulant.

 In our simulations,  the behaviour of susceptility of imaginary part of Polayakov loop ${\rm Im}(L)$ at $am=0.024$ can give us clear signal to determine the nature of transition, while at other quark mass, it is difficult to determine the nature of transition.

 The values of $B_4(\beta_c,\infty)$ extracted through  fitting procedure are not in consistent with what are  expected.  This is because logarithmic scaling corrections will be present near the tricritical point~\cite{ls,deForcrand:2010he},  and our  simulations are carried out on finite size volume  on which large finite size corrections are observed  in simpler spin model~\cite{Billoire:1992ke}. However, the critical exponent $\nu$ is not sensitive to finite size corrections~\cite{deForcrand:2010he}. So index $\nu$  extracted through  fitting procedure  can provide us information to
 determine the transition nature.

 In our simulation,  we can find that the values of $B_4$ on lattice with spatial volumes $12^3,$
 $16^3,$ $20^3$ intersect approximately at one point at quark masses $am=0.024, 0.026,0.050$, while at other quark mass, it is difficult to find intersection point for $B_4$'s from three spatial volumes. It is expedient to determine the intersection point from two spatial volumes as indicated in Table.~\ref{critical_beta_B4}.

 Taking what mentioned above into account, further  work along this direction which can  provide crosscheck is expected, especially simulations with larger time extent which is being under our consideration.

\begin{acknowledgments}
We thank Philippe de Forcrand for valuable helps.
 We modify the MILC collaboration's
public code~\cite{Milc} to simulate the theory at imaginary chemical
potential.  We use the fortran-90 based multi-precision software~\cite{fortran}.   This work is supported by
the National Science Foundation of China (NSFC)  under Grant Nos.~(11347029). The work was carried out at National Supercomputer Center in Wuxi, We appreciate  the help of Qiong Wang and Zhao liu when carrying out the computation.
\end{acknowledgments}

\end{document}